# Random Fiber Laser

Christiano J. S. de Matos[1,*], Leonardo de S. Menezes[2], Antônio M. Brito-Silva[3], M. A. Martinez Gámez[4], Anderson S. L. Gomes[2], and Cid B. de Araújo[2]

1– Programa de Pós-Graduação em Engenharia Elétrica, Universidade Presbiteriana Mackenzie, São Paulo-SP, 01302-970, Brazil

2 – Departamento de Física, Universidade Federal de Pernambuco, Recife-PE, 50670-901, Brazil

3 – Programa de Pós-Graduação em Ciência de Materiais, Universidade Federal de Pernambuco, Recife-PE, 50670-901 Brazil

4 – Centro de Investigaciones en Optica, C.P. 37150 Leon, Gto., Mexico

* – Corresponding autor. E-mail: cjsdematos@mackenzie.br

**Abstract**

We investigate the effects of two dimensional confinement on the lasing properties of a classical random laser system operating in the incoherent feedback (diffusive) regime. A suspension of 250nm rutile ($TiO_2$) particles in a Rhodamine 6G solution was inserted into the hollow core of a photonic crystal fiber (PCF) generating the first random fiber laser and a novel quasi-one-dimensional RL geometry. Comparison with similar systems in bulk format shows that the random fiber laser presents an efficiency that is at least two orders of magnitude higher.

**PACS Codes:** 42.55.Mv Dye lasers; 42.55.Wd Fiber lasers; 42.55.Zz Random lasers; 42.55.Tv Photonic crystal lasers and coherent effects

The emission of laser-like radiation in highly scattering gain media [1,2], which became known as random lasers (RL), has received considerable attention for over a decade due to its unique properties and to a number of potential applications [2]. The optical feedback in RL is provided within the amplifying medium by random light paths that result in increased emission intensity and reduced emission linewidth. The RL action can take place in two distinct regimes [2]. When the length of the scattering gain medium is greater than the photon mean free path, and this path, in turn, is greater than the emission wavelength, light propagation is diffusive. In this case, the probabilistic nature of diffusion means that interference contributes negligibly to the feedback process which is incoherent. On the other hand, if the photon mean free path and the emission wavelength have the same order of magnitude, localization of the radiation field occurs within the structure and coherent feedback is possible [3].

An important concept for RLs is that of the $\beta$ factor, defined (both for conventional and random lasers) as the fraction of spontaneous emission that seeds the laser process [4]. The value of $\beta$ is directly related to the sharpness of the laser threshold, with a thresholdless laser having $\beta = 1$. While in conventional lasers $\beta$ depends on both the fraction of emitted photons that is collected by the cavity mirrors and the fraction of the spontaneous emission spectrum that overlaps with the laser spectrum, in RLs $\beta$ depends only on the latter. Thus, $\beta$ in RLs is usually considerably larger. However, the lack of directionality in these devices has largely limited their application as high $\beta$ sources.

Recently, a system consisting of a stack of reflecting glass slides of variable thicknesses alternated with amplifying layers was presented [5]. This system was shown to behave as a one-dimensional random laser in the strong localization regime, in which pump and emission longitudinal modes arise deep in the sample. When these two modes overlap with each other and with the amplifying layers, a very low RL threshold is obtained and the RL action is very efficient. Such an approach, however, is not applicable to the diffusive (incoherent feedback) RL regime, in which stable longitudinal modes are never formed. It is worth noting that a theoretical model for 1D RLs operating in the strong localization regime was also reported [6].

A configuration that is extensively used for random lasing with incoherent feedback was demonstrated by Lawandy et al. [1], in which scattering sub-wavelength particles are suspended in a dye solution. Pump scattering restricts population inversion to a thin disk close to the solution surface [7]. This means that photons can diffuse out of the active region, contributing to loss and to an increase in lasing threshold. Decreasing this threshold generally requires an increase in dye and/or scatterer concentration. Alternative methods have been described that somewhat are able to increase the lasing efficiency. In one of these [8], external feedback is provided by a mirror that increases the photon lifetime in the solution and, thus, decreases the threshold pump energy. In another approach [9], surface plasmon excitation is exploited to locally increase the radiation field and to reduce the lasing threshold. In addition to presenting an improved efficiency, the latter scheme exhibits some emission directionality, which is not observed in conventional RL.

Both directionality and increased efficiency can be expected if the scattering gain medium is inside a waveguide that transversely confines light. A 1-mm-thick planar waveguide RL where liquid crystal droplets acted as scatterers was recently demonstrated [10]. When an intense electric field was applied perpendicularly to the sample's plane, transverse scattering was inhibited. Light was scattered along the plane and transversely confined by reflection at the boundaries of the planar geometry. As a consequence, a reduction in the pump intensity threshold and a decrease in the emission linewidth by 33% were observed. It is clear that a channel waveguide, providing two-dimensional confinement, would further improve the RL characteristic. However, such a configuration has not yet been investigated for RLs in any regime, possibly due to the fact that a practical cladding material, presenting a refractive index that is lower than that of the scattering gain solution, was unavailable.

The advent of hollow-core photonic crystal fibers (PCFs) [11] opens new possibilities. In PCFs the microstructured cladding presents an effective refractive index that is a weighted average between the indices of glass and air and may have values below 1.3. In addition, hollow-core PCFs may present photonic-bandgap guidance, which confines light to the core irrespective of its refractive index. These features have enabled the design and fabrication of liquid-core PCFs and a large number of theoretical and experimental studies have already been reported for various applications [12,13]. Of particular importance to the present letter is the demonstration of fiber dye lasers [13], obtained when PCFs or capillary fibers were filled with a dye solution and end-pumped to yield conventional laser action.

In this letter, we demonstrate random laser action in a channel waveguide geometry consisting of a hollow-core PCF. The PCF (Crystal Fibre A/S, model HC-1550-02) had core diameter, cladding pitch and cladding air-filling fraction of ~10.9 μm, 3.8 μm and >90%, respectively, and its cross-sectional profile is depicted in the inset of Fig. 1. The scattering gain medium, consisting of a suspension of 250-nm rutile ($TiO_2$) in a solution of Rhodamine 6G in ethylene glycol, was selectively inserted into the core, leaving the cladding microstructure filled with air. The dye and scatterer concentrations were varied in the range $10^{-5}$ M $\leq \rho_{dye} \leq 10^{-3}$ M and $10^7$ cm$^{-3}$ $\leq \rho_{scatt} \leq 10^9$ cm$^{-3}$, respectively. For comparison, Rhodamine 6G solutions without rutile particles were also investigated.

As the cladding effective refractive index was numerically estimated to be 1.27 and the refractive index of ethylene glycol is ~1.43, the PCF core was able to confine and guide light through total internal reflection. For $\rho_{scatt} = 10^8$ cm$^{-3}$ one can calculate that the average distance between scatterers is 21 μm. As the mean free path of photons within the suspension is expected to be a few times larger than this average distance, scattering events are unlikely to occur for photons that radially cross the PCF core. Consequently, total internal reflection is the main mechanism transversely confining light within the gain medium. Scattering events mainly occur along the fiber axis, which results, to the best of our knowledge, in the first quasi-one-dimensional (1D) RL reported in the diffusive (incoherent feedback) regime. Also, we note that transverse confinement is not obtained in the 1D RL presented in [5].

To fill the core of the PCF with the gain medium, while leaving the cladding holes filled with air, we first used a commercial fiber splicer to close the holes around the hollow core following the procedure described in [14,15]. To insert the solution into the PCF core we removed the metallic cannula of a 3 mℓ syringe needle from its plastic support. Then, we introduced ~5-10 mm of a ~50-mm-long piece of PCF into the hole of the plastic support, sealing this hole with a UV curable glue. Next, we attached the plastic support to the syringe, filling the latter with ~1.0 mℓ of the liquid gain medium. A simple setup was used to press the syringe's plunger, which was kept under pressure during the entire process. This guaranteed that no air bubbles were formed in the PCF core. After initiating the PCF core filling process, it took approximately one minute to observe a droplet appearing at the end of the PCF.

The experimental setup for the optical measurements is shown in Fig. 1. The second harmonic (wavelength: 532 nm) of a Q-switched Nd:YAG laser (7 ns, 5 Hz) was used to laterally pump the PCF. An iris with a 4 mm diameter aperture is used for obtaining a top-hat beam which is focused by a 50 mm focal length cylindrical lens on the final portion of the PCF, so that a 50 μm × 4 mm area of the PCF receives nearly uniform illumination. An advantage of this configuration over its bulk counterparts is that, due to the small core diameter, all the dye molecules within the 4-mm section are equally excited, and not a thin superficial layer as in previous schemes. The maximum intensity (fluence) inside the PCF core was 144 MW/cm$^2$ (2.1 J/cm$^2$). The light emitted along the PCF axis was then collected by a 0.44 numerical aperture optics and sent to a spectrometer. Care was taken to align the cylindrical lens with the PCF's axis, since it was observed that the recorded spectra substantially vary with this alignment. We also controlled the laser intensity to prevent bleaching of the dye molecules even when the highest optical intensities were used.

Figure 2 shows plots of the emission spectra for various pump intensities and for different configurations. Figure 2(a) shows results for the conventional RL scheme for a suspension with $\rho_{dye} = 10^{-4}$ M and $\rho_{scatt} = 10^{8}$ cm$^{-3}$ placed inside a cylindrical quartz cuvette (radius = height = 10 mm). For a fair comparison with results obtained with the random fiber laser setup, pumping in the present case also occurred along a 4-mm line. The emission linewidth remains almost constant when the pump intensity is increased from 9 to 144 MW/cm$^2$. The not so broad linewidths measured (~22 nm FWHM), as compared to the results reported in the literature [1,9,16], occur because the optics at the fiber output is optimized to collect light from the pumped region and not from the unpumped surrounding, where photon absorption and re-emission broadens the linewidth. In Fig. 2(b), the spectra are shown for a medium with $\rho_{dye} = 10^{-4}$ M and $\rho_{scatt} = 0$ cm$^{-3}$, placed inside the PCF core. Again, it is observed that the emission spectral linewidth do not change much for the same pump intensity range, but in this case the spectra are narrower than in the previous situation. This is due to the reduced transverse dimensions inside the PCF, as compared to the bulk case, that further prevents the absorption and emission of Stokes photons with smaller energies that contribute for broadening of the emission spectrum.

Finally, Fig. 2(c) shows the results of the experiment with rutile particles inside the PCF. The solution into the core had $\rho_{dye} = 10^{-4}$ M and $\rho_{scatt} = 10^{8}$ cm$^{-3}$. The linewidth significantly decreases when the pump intensity increases. The results demonstrate that the presence of scatterers is essential for observing spectral line narrowing. The emission red-shift as the pump intensity is increased has been previously observed [4] and described [17]. In the present case, for low pump intensities spontaneous emission dominates and detected photons are mostly emitted from the region closest to the fiber output. As pump intensity is increased the higher gain allows for photons emitted further into the fiber to reach the output. The long path within the gain medium then increases photon absorption and re-emission, causing the red-shift [17].

In order to better characterize the laser-like behavior of the last system, we plotted in Fig. 3(a) and Fig. 3(b) the emission peak intensity and spectral linewidth (FWHM), respectively, as functions of the pump intensity. A large increase in the emission peak intensity is observed for intensities larger than the threshold pump intensity ($I_{thr}$ ~40 MW/cm$^{2}$). Unlike bulk RLs, the random fiber laser presented a threshold behavior even if the emission intensity was integrated over its spectrum, as shown as open symbols in Fig. 3(a). As pointed in Ref. [4], such a behavior indicates that the threshold is not only a result of spectral narrowing, but also of the existence of directionality in the emission above threshold. The observed threshold behavior is, therefore, evidence that the emission in the random fiber laser is directional. Figure 3a also shows that when rutile was not present in the PCF core (circles) no intensity threshold behavior could be detected. Figure 3(b) shows that at threshold the emission linewidth reduces from ~24 nm to ~7 nm. This behavior is typical of RL action. Figure 3(b) also shows the linewidth dependence on pump intensity when a Rhodamine solution (without scatters) is inside the PCF. Note that no laser threshold is observed. The slight reduction in linewidth, which has also been observed in [1], is believed to be the effect of the spectral dependence of gain, which for high gain values naturally causes a narrowing of the amplified spontaneous emission, especially in long gain media. This behavior is routinely observed in active optical fibers.

It is important to compare our results with those reported for other RL schemes. For this purpose, we define a figure of merit FOM = $(I_{thr} \cdot \rho_{dye} \cdot \rho_{scatt})^{-1}$ where $I_{thr}$ is in MW/cm$^2$, $\rho_{dye}$ is the gain medium concentration in mol/$\ell$ and $\rho_{scatt}$ is the density of scatterers in cm$^{-3}$. Lawandy *et al.* [1] investigated the classical RL system: a suspension rutile particles (diameter: 250 nm) in an alcoholic solution of Rhodamine 640, obtaining FOM = $1.3 \times 10^{-8}$ cm$^5 \ell$/MW.mol. In Ref. [18], the authors investigate the temporal and spectral behavior of a RL for different values of $\rho_{dye}$ and $\rho_{scatt}$, obtaining at best FOM = $1.1 \times 10^{-9}$ cm$^5 \ell$/MW.mol. RL systems in which both the dye molecules and the scattering particles were imbedded in a polymeric host were also studied [16] and the results lead to FOM = $1.7 \times 10^{-9}$ cm$^5 \ell$/MW.mol. The use of external feedback to decrease the pump intensity threshold [8] was investigated in a suspension of TiO$_2$ in a Rhodamine 640 solution, with results leading to FOM ~ $3.5 \times 10^{-8}$ cm$^5 \ell$/MW.mol. In Ref. [19], the authors investigated RL action of a dye solution containing aggregates of ~700 nm in diameter formed by ~100 nm in length TiO$_2$ rods. In this case, the concentration of scattering particles is not given, but assuming the same concentration as in the present paper, one gets FOM = $4.4 \times 10^{-8}$ cm$^5 \ell$/MW.mol (this is an approximate value because the local field enhancement contributes to lowering the RL threshold in fractal structures). More recently [20], another RL scheme, using 55 nm in diameter silver nanoparticles instead of TiO$_2$ particles as scattering centers, was introduced having FOM = $6.1 \times 10^{-12}$ cm$^5\ell$/MW.mol. In the present work, we obtain FOM = $2.5 \times 10^{-6}$ cm$^5\ell$/MW.mol. The significantly higher efficiency obtained here in comparison with the previously reported results is due to the novel channel waveguide RL geometry, that favors a transverse feedback mechanism contributing to an increase of the photon lifetime within the scattering gain medium. A comparison with the 1D RL studied in [5] is not straightforward due to the substantial differences in the construction of their system. Nevertheless, assuming $\rho_{dye}$ in their case to be the cube of the average distance between reflecting surfaces, one obtains FOM = $2.9 \times 10^{-3}$ cm$^5\ell$/MW.mol. This value is substantially higher than that of the present work, but is achieved due to a particular feature (namely, the emergence of longitudinal modes) of RLs in the strong localization regime. In contrast, the channel waveguide RL approach is applicable to RLs in any regime and is of particular importance for incoherent feedback RLs.

The transverse feedback results in peculiar characteristics that are not present in other RL configurations. First, directionality, which is an attractive feature for some potential applications, is naturally obtained. Second, although a specific characterization was not performed, the multiple lateral reflections are expected to generate transverse laser modes. This feature is a direct consequence of the studied configuration not being a RL in the transverse direction. Using the ethylene glycol refractive index and the effective cladding index, the liquid-core PCF numerical aperture is calculated to be ~0.657, which from simple waveguide estimates [21] results in over 700 guided modes. In a ray optics approximation, the ray corresponding to the highest-order guided mode makes a 27° angle with the fiber axis. Note, nevertheless, that the scattering events are expected to result in a random coupling between the guided modes.

It is possible to derive an estimate for $\beta$ in the random fiber laser. Were it a bulk RL, the method used in [4] could be directly employed. However, the fact that directionality is present means that a modification is necessary. Here we estimate $\beta$ as in [4] and subsequently multiply it by the fraction of isotropically emitted photons that are collected by the fiber core. We obtain $\beta = 1.4\times10^{-2}$, which is 10 times smaller than typical values for bulk RLs but is still significantly higher than those for most conventional lasers [4]. The demonstrated device, therefore, presents a compromise between a high beta value and directional emission, occupying a so far unexplored region in the laser-parameter space.

In conclusion, we reported the operation of a quasi 1D RL in the diffusive regime, using a PCF as a confining waveguide. Due to the light confinement properties of this structure, the RL action is more efficient than that obtained in similar RLs in bulk format previously reported in the literature. For the fully understanding this system a theoretical model, which also considers the possibility of obtaining single-mode random lasing in a fiber geometry, will be presented in the near future.

**Acknowledgments**

The financial support from the Brazilian agencies Conselho Nacional de Desenvolvimento Científico e Tecnológico (CNPq) and Coordenação de Aperfeiçoamento de Pessoal de Ensino Superior (CAPES/PROCAD), and from the Mexican agency Consejo Nacional de Ciencia y Tecnologia (CONACYT under grant No. 47182) are acknowledged. The PCF's scanning electron micrograph was provided by Crystal Fibre A/S.

**References**

1. N. M. Lawandy *et al.*, Nature **368**, 436 (1994).

2. H. Cao, Waves in Random Media **13**, R1 (2003).

3. H. Cao *et al.*, Phys. Rev. Lett **82**, 2278 (1999).

4. G. van Soest and A. Lagendijk, Phys. Rev. E **65**, 047601 (2002).

5. V. Milner and A. Z. Genack, Phys. Rev. Lett. **94,** 073901 (2005).

6. A. L. Burin *et al.*, Phys. Rev. Lett. **88**, 093904 (2002).

7. D. Wiersma *et al.*, Nature **373**, 203 (1995).

8. P. C. de Oliveira *et al.*, Opt. Lett. **22,** 895 (1997).

9. M. A. R. C. Alencar *et al.*, J. Opt. Soc. Am. B **20**, 564 (2003).

10. S. Gottardo *et al.*, Phys. Rev. Lett. **93**, 263901 (2004).

11. J. C. Knight, Nature **424**, 847 (2003).

12. J. M. Fini, Meas. Sci. Technol. **15**, 1120 (2004); Y. Huang *et al.*, Appl. Phys. Lett. **85**, 5182 (2004); S. Yiou *et al.*, Opt. Express **13**, 4786 (2005); F. Du *et al.*, Appl. Phys. Lett. **85**, 2181 (2004); C. M. B. Cordeiro *et al.* Meas. Sci. Technol. (to be published).

13. A. E. Vasdekis *et al.*, Opt. Express **15**, 3962 (2007).

14. L. Xiao *et al.*, Opt. Express **13,** 9014 (2005).

15. C. M. B. Cordeiro *et al.*, Opt. Express **14,** 8403 (2006).

16. R. M. Balachandran *et al.*, Appl. Opt. **35,** 640 (1996).

17. K. Totsuka *et al.* Phys. Rev. B **59**, 50 (1999).

18. W. L. Sha *et al.*, Opt. Lett. **19,** 1922 (1994).

19. H. Z. Wang *et al.*, Opt. Lett. **23,** 777 (1998).

20. G. D. Dice *et al.*, Appl. Phys. Lett. **86**, 131105 (2005).

21. A. W. Snyder and J. D. Love, *Optical Waveguide Theory* (Chapman & Hall, London, 1983).

**Figure Captions**

Fig. 1: (color online) Side view of the experimental setup. VNDF: variable neutral density filter; CL: 50 mm focal length cylindrical lens; SH: sample holder; PCF: photonic crystal fiber; SL: 50 mm focal length spherical lens; SPEC: spectrometer. Inset: Scanning electron micrograph of the photonic crystal fiber used in the experiment.

Fig. 2: (color online) Emission spectra measured along the excitation region for different pump intensities and different configurations. (a) A $10^{-4}$ M solution of Rhodamine 6G in ethylene glycol containing $10^8$ cm$^{-3}$ rutile particles inside a cell. (b) A $10^{-4}$ M solution of Rhodamine 6G in ethylene glycol inside the photonic crystal fiber, without scattering particles. (c) A $10^{-4}$ M solution of Rhodamine 6G in ethylene glycol inside the photonic crystal fiber with $10^8$ cm$^{-3}$ rutile particles.

Fig. 3: (color online) (a) Emission peak (solid symbols) and spectrum-integrated (open symbols) intensities for $10^{-4}$ M Rhodamine in ethylene glycol with (squares) and without (circles) $10^8$ cm$^{-3}$ rutile particles in the core of the PCF. (b) Emission linewidth (FWHM) for the same configurations. Symbols represent the same conditions as in (a). The solid lines are guides to the eye.

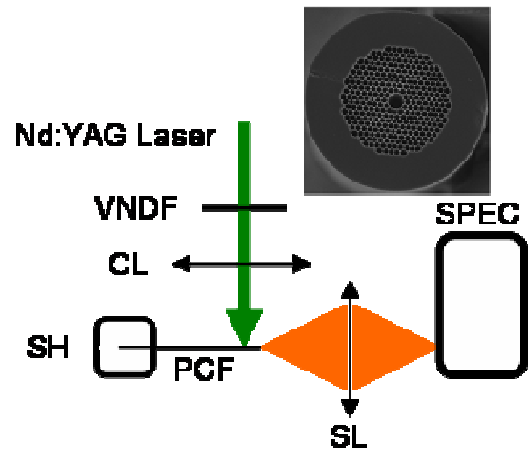

Fig. 1 – de Matos *et al.*

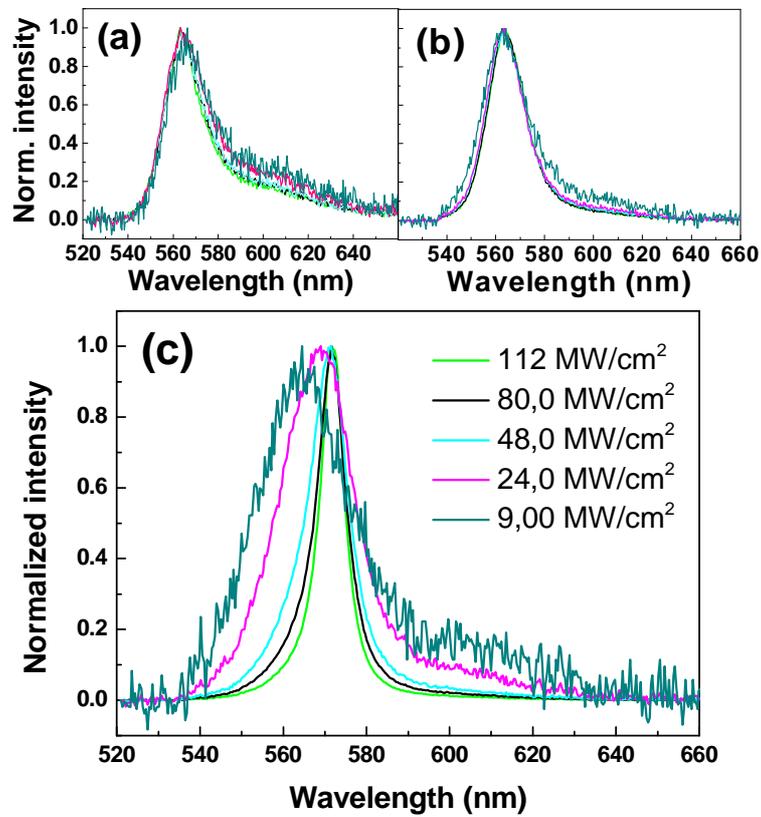

Fig. 2 – de Matos *et al.*

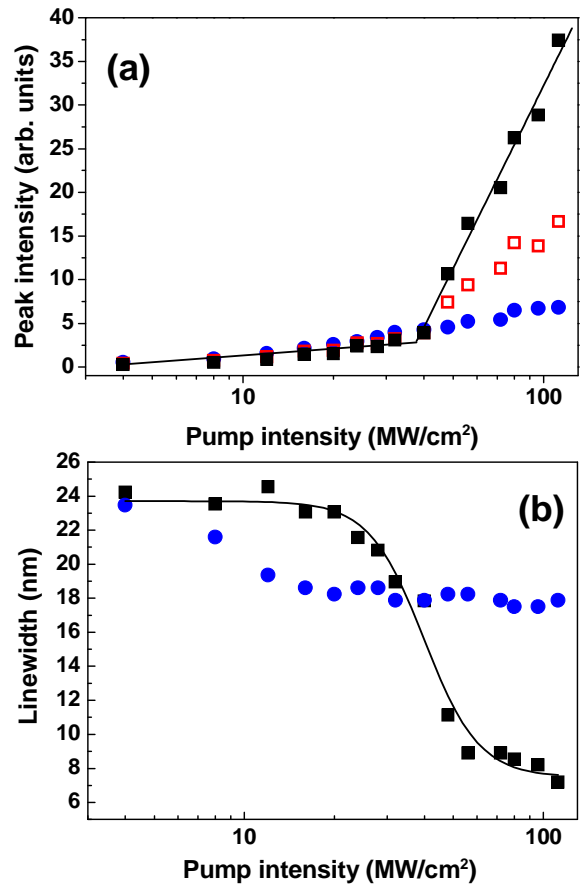

Fig. 3 – de Matos *et al.*

15